\documentclass[namedreferences]{SolarPhysics}
\usepackage[optionalrh]{spr-sola-addons} 
\usepackage{graphicx}        
\usepackage{url}             

\newcommand{\Imax}{IMaX}
\newcommand{\Islid}{ISLiD}
\newcommand{\Cosm}{CoSM}
\newcommand{\Cwcom}{CW-Com}
\newcommand{\Cwao}{CW-AO}%
\begin{document}

\begin{article}

\begin{opening}
\title{The Wave-Front Correction System for the {\it Sunrise} Balloon-Borne Solar Observatory}


\author{T.~\surname{Berkefeld}$^{1}$\sep
        W.~\surname{Schmidt}$^{1}$\sep
        D.~\surname{Soltau}$^{1}$\sep
        A.~\surname{Bell}$^{1}$\sep
        H.P.~\surname{Doerr}$^{1}$\sep
        B.~\surname{Feger}$^{1}$\sep
        R.~\surname{Friedlein}$^{1}$\sep
        K.~\surname{Gerber}$^{1}$\sep 
        F.~\surname{Heidecke}$^{1}$\sep
        T.~\surname{Kentischer}$^{1}$\sep
        O.~\surname{v. d. L\"uhe}$^{1}$\sep 
        M.~\surname{Sigwarth}$^{1}$\sep
        E.~\surname{W\"alde}$^{1}$\sep
        P.~\surname{Barthol}$^{2}$\sep
        W.~\surname{Deutsch}$^{2}$\sep
        A.~\surname{Gandorfer}$^{2}$\sep
        D.~\surname{Germerott}$^{2}$\sep
        B.~\surname{Grauf}$^{2}$\sep
        R.~\surname{Meller}$^{2}$\sep
        A.~\surname{\'Alvarez-Herrero}$^{6}$\sep
        M.~\surname{Kn\"olker}$^{4}$\sep
        V.~\surname{Mart\'\i nez Pillet}$^{3}$\sep
        S.K.~\surname{Solanki}$^{2}$\sep
        A.M.~\surname{Title}$^{5}$
  }
    \institute{
    $^{1}$ Kiepenheuer-Institut f\"{u}r Sonnenphysik, Sch\"{o}neckstra\ss e 6, D-79104 Freiburg, Germany 
     email: \url{berke@kis.uni-freiburg.de} \\
    $^{2}$ Max-Planck-Institut f\"{u}r Sonnensystemforschung, Max-Planck-Stra\ss e 2, D-37191 Katlenburg-Lindau, Germany\\
    $^{3}$ Instituto de Astrof\'{i}sica de Canarias, C/ Via L\'{a}ctea, s/n, E38205 - La Laguna (Tenerife), Spain \\
    $^{4}$ High Altitude Observatory\thanks{HAO/NCAR is sponsored by the National Science Foundation}, P.O. Box 3000, Colorado 80301, USA\\
    $^{5}$ Lockheed Martin Solar and Astrophysics Laboratory, Bldg. 252, 3251 Hanover Street, Palo Alto, CA 94304, USA \\
    $^{6}$ Instituto Nacional de T\'{e}cnica Aeroespacial, E-28850, Torrej\'{o}n de Ardoz, Madrid, Spain\\
}





\date{\today}


\abstract
{}
{This paper describes the wave-front correction system developed
  for the {\it Sunrise} balloon telescope, and provides information about its
  in-flight performance.}
{For the correction of low-order aberrations, a Correlating Wave-Front Sensor
  (CWS) was used. It consisted of a six-element Shack-Hartmann wave-front sensor
  (WFS), a fast tip-tilt mirror for the compensation of image motion, and an
  active telescope secondary mirror for focus correction.}
{The CWS delivered a stabilized image with a precision of 0.04~arcsec (rms),
  whenever the coarse pointing was better than $\pm$~45~arcsec peak-to-peak. The automatic
  focus adjustment maintained a focus stability of 0.01~waves in the focal
  plane of the CWS. During the 5.5~day flight, good image quality and stability
  was achieved during 33~hours, containing 45~sequences that lasted between 10
  and 45~minutes.}
{}

}
\runningtitle{Sunrise Wave-front correction system}
\runningauthor{Berkefeld {\it et al.}}
\end{opening}

\section{Introduction}

\medskip

The {\it Sunrise}\footnote{{\it Sunrise} is an international collaboration of the
  Max-Planck Institut f\"ur Sonnensystemforschung (Katlenburg-Lindau, Germany),
  the High Altitude Observatory (Boulder, U.S.), the Lockheed-Martin
  Solar and Astrophysics Laboratory (Palo Alto, U.S.), the \Imax{} Consortium
  (Tenerife, Granada, Madrid, Spain) and the Kiepenheuer-Institut f\"ur
  Sonnenphysik (Freiburg, Germany).} balloon-borne telescope for solar
observations performed its first science flight in June 2009 on a NASA
long-duration balloon flight from Kiruna, Sweden to Somerset Island in
North-East Canada (see Barthol {\it et al.}, 2010, Schmidt {\it et al.}, 2010). After a perfectly smooth launch,
{\it Sunrise} headed West, across the Northern Atlantic and Greenland. About
1.8~TByte of science data were collected. A few days after landing, the data
disks, the telescope and the gondola were safely recovered. The processing of
the scientific and engineering data is underway, as well as an inspection of the
flight hardware.

Stratospheric balloon-borne telescopes have two fundamental advantages over
ground based telescopes: they permit UV observations and they provide a
seeing-free image quality over the full field of view
(FoV). However, pointing to the Sun and tracking a feature on the solar surface
is a formidable task, especially for a telescope hanging under a balloon that is
driven by stratospheric winds at an altitude of 36~km. In addition to the
apparent (diurnal and seasonal) motion of the Sun, there is a number of
oscillatory modes that may be induced by variable winds in the stratosphere, and
taken up by the balloon-gondola system. The tip-tilt correction of CWS was built
with a range of $\pm$~45~arcsec, and a closed-loop bandwidth of 60~Hz (6~db attenuation of tip-tilt). The
instrument worked reliable throughout the flight. Closed-loop operations were
possible, whenever the gondola pointing was within the angular range of the CWS
tip-tilt mirror.

Section~\ref{sec:od} explains the optical design, Section~\ref{sec:wfc} deals with the wave-front
correction algorithms. The electronics and tip-tilt hardware is described in Section~\ref{sec:hardware}. The control software 
is addressed in Section~\ref{sec:sw}, followed by a discussion of the
in-flight performance in Section~\ref{sec:perf}. Finally, Section~\ref{sec:concl}
concludes this paper.
   

\section{Opto-Mechanical Design}
\label{sec:od}
\begin{figure}
  \resizebox{\hsize}{!}{\includegraphics{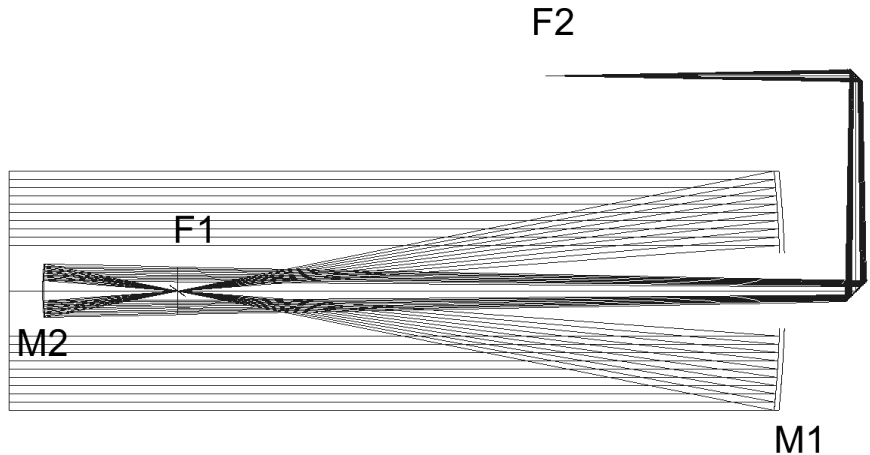}}
  \caption{Optical design of the {\it Sunrise} telescope.}
  \label{f:optical_design}
\end{figure}

{\it Sunrise} is a 1~m Gregory-type telescope (see Figure~\ref{f:optical_design} and Barthol {\it et al.} (2010) for a detailed description). 
The parabolic primary mirror M1 (f/2.42)
provides a full disk image in its focal plane. Here a field stop which is also a
heat rejection device reflects 99~\% of the sunlight. Only a field of view of 200 arcsec 
passes to the elliptical secondary mirror M2 which delivers a f/24 focus to F2.

{\it Sunrise} feeds two scientific instruments: SuFI (Sunrise Filter Imager,  see Gandorfer {\it et al.}, 2010), an
UV-imager, and \Imax{} (Imaging Magnetograph Experiment, see Mart\'\i nez Pillet {\it et al.}, 2010), a compact filter based
magnetograph. In addition, the CWS (Correlating Wave-Front Sensor) is fed by
{\it Sunrise} to measure the image shift and possible low order aberrations (focus
and coma), which are due to misalignment of M1 or M2 caused by thermo-elastic deformations of the 
telescope during flight. 

\begin{figure}
  \resizebox{100mm}{!}{\includegraphics{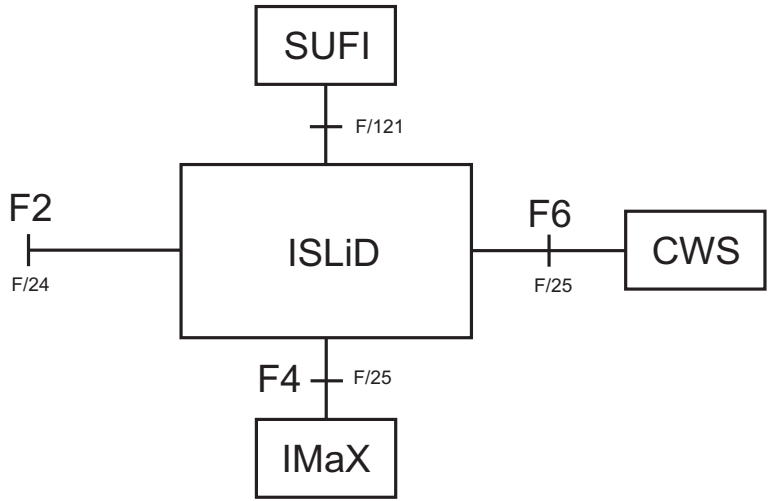}}
  \caption{Scheme of the Image Stabilization and Light Distribution 
    system (\Islid{}).}
  \label{f:islid}
\end{figure}

For tip-tilt correction, a pupil image of appropriate size is needed where the
tip-tilt mirror can be placed. The mirror size determines the necessary mirror
stroke (the smaller the mirror the larger the stroke). On the other hand a
larger pupil means longer focal length. The optical solution which was chosen is
a field lens which provides a 35 mm pupil image. Here the tip-tilt mirror M105
(notation compatible with official {\it Sunrise} documentation) was placed. This
mirror is part of the \Islid{} (Image Stabilization and Light
Distribution) system (see Gandorfer {\it et al.}, 2010) and is further described in Sect.~\ref{sec:hardware}. 
While maintaining the diffraction-limited image quality, \Islid{}
provides the following:
\begin{itemize}
\item the desired f/121 image for the UV channel (SuFI)
\item an f/25 image for \Imax{} (visible)
\item an f/25 image to the CWS (visible)
\item a possibility to feed a third instrument (near-IR spectrograph, planned for the second flight)
\end{itemize}
The main functions of \Islid{} are shown schematically in Figure~\ref{f:islid}: For this purpose, a modified
Schwarzschild system is used to magnify the image for SuFI. Four relay lenses
demagnify the image and shift the pupil so that \Imax{} and the CWS are
illuminated with an f/25 beam, with a pupil at infinity. An additional correction
element accounts for the beam splitter not used in collimated light. 

\begin{figure}
  \resizebox{\hsize}{!}{\includegraphics{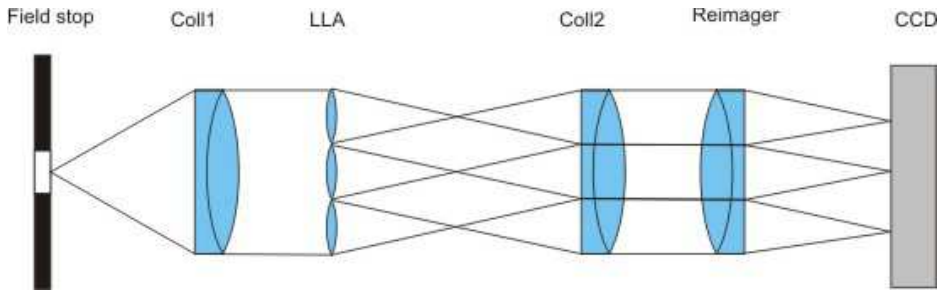}}
  \caption{Scheme of the wave-front sensor of {\it Sunrise}. The field stop coincides
    with a focal plane of the telescope.}
  \label{f:wfs}
\end{figure}

\begin{figure}
  \resizebox{80mm}{!}{\includegraphics{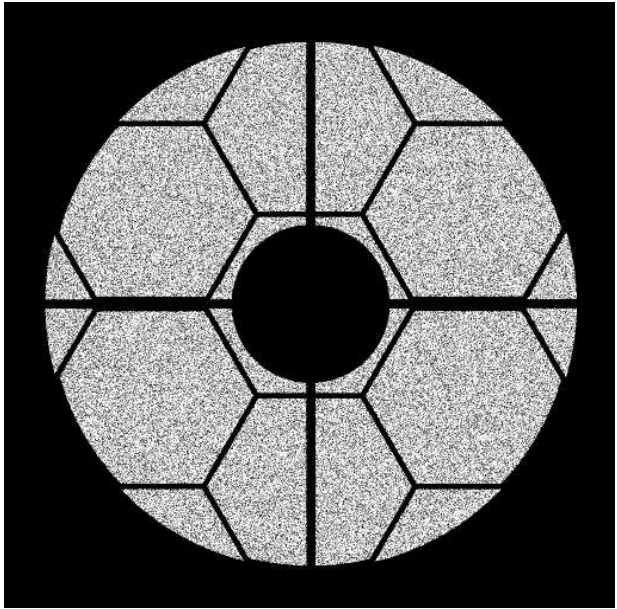}}
  \caption{Illumination pattern of the CWS lenslet array. The image of the 1~m
    entrance pupil provides a homogeneous illumination of the six peripheral
    micro-lenses, except for the (small) influence of the spiders. The central
    lenslet is obscured by the secondary mirror, and is not used.}
  \label{f:lla}
\end{figure}

The CWS is a Shack-Hartmann (SH) type wave-front sensor working at 500~nm. Its optical design is
shown in Figure~\ref{f:wfs}. It picks up the image at the entrance focus F6 which also defines the system focus of {\it Sunrise}. 
Differential foci between the science instruments and the CWS are not corrected by the CWS and have been avoided by a careful system alignment.
A field stop with a square area of $1.54 \times 1.54$ mm$^2$ limits the field of view to 13~arcsec. Coll1 (f =
100~mm) images a 4.6~mm pupil on a hexagonal lenslet array (LLA, f = 81~mm)
which consists of 7 hexagonal micro-lenses. The geometry of the lenslet array is
shown in Figure~\ref{f:lla}. The central one is obscured by the shadow of the
telescope's secondary. A relay lens system consisting of a collimator Coll2 (f = 160~mm)
and an imaging lens (f = 90~mm) picks up the focal plane of the LLA and images
the subfields on the wave-front sensor camera. The image scale is chosen so that
1~arcsec corresponds to 5~pixels in the camera plane. 
The left panel of
Figure~\ref{im:imwfs} in Section~\ref{sec:perf} shows the six images on the
high-speed camera. It is an original flight picture taken as a screen copy from
the ground control system. Solar granulation is clearly visible at high contrast
and good image sharpness. 


\section{Wave-Front Correction}
\label{sec:wfc}
\subsection{Overview}
\begin{figure*}
  \resizebox{\hsize}{!}{\includegraphics[width=16.5cm,angle=0]{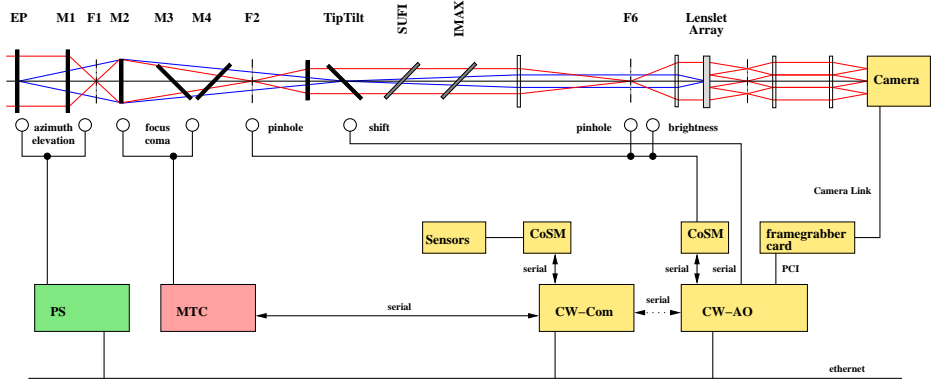}}
  \caption{Scheme of the {\it Sunrise} CWS hardware and communication lines. CWS electronics 
    components are shown as yellow boxes, PS denotes the gondola pointing system and MTC the main telescope controller.}
  \label{fig:overview}
\end{figure*}

An overview of the CWS architecture is given in Figure~\ref{fig:overview} where
black lines are communication paths that are typically labeled with the interface
used.  The upper region shows the light path from the entrance pupil (EP) of the
telescope to the camera. CWS electronics components  
are shown as yellow boxes. In addition the figure
shows the pointing system (PS, green box) provided by the High Altitude Observatory in
Boulder, U.S., and the main telescope controller (MTC, pink box) delivered by the
manufacturer of the telescope, Kayser Threde in Munich, Germany. The actual
wave-front correction is controlled by a computer named \Cwao{} depicted in the right
part of the overview.

The main components of the wave-front correction system are the following
devices: the Shack-Hartmann wave-front sensor (Section~\ref{sec:od}), a high-speed
camera (Section~\ref{sec:camera}), the control loop computer that converts the
shifts measured with the SH sensor into actuator signals (Section~\ref{sec:meu}),
the fast tip-tilt mirror (1700~Hz update rate, Section~\ref{sec:tt}) and the slow
secondary mirror M2 (0.1~Hz update rate) (see Schmidt {\it et al.} 2004, 2006).
The telescope secondary, M2, has tight alignment tolerances with respect to M1,
therefore M2 is mounted on a motorized {\it xyz} stage that is controlled by the CWS
to ensure proper alignment. Lateral ({\it xy}) misalignment causes coma and image
shift, while axial ({\it z}) misalignment causes spherical aberrations (negligible)
and defocus.

\subsection{Algorithms}
The Shack-Hartmann wave-front reconstruction is done in much the same way as in
the solar AO system KAOS at the Vacuum Tower Telescope on Tenerife (see von der L\"uhe {\it et al.}, 2003).

The data reduction consists of the following steps for each 64$\times$64 pixel
sub-aperture, in the order given below:
\begin{itemize}
\item dark/flat correction
\item subtraction of the average (of this sub-aperture) intensity
\item removal of the  intensity gradient across the FoV
\item application of a Hamming window $W(x,y) = w(x)\cdot w(y)$ with $w(i) =
  0.54+(0.54-1)\cdot {\rm cos}(2\pi\cdot i/(64-1))$
\item addition of the former average intensity
\item intensity normalization
\end{itemize}
The subtraction and addition of the average intensity is advisable for the
application of the Hamming window which eliminates aliasing problems of the
FFT-based cross correlation. The removal of the intensity gradients and the
intensity normalization are required to get the correct position of the
correlation peak.

The reduced sub-aperture images are now cross correlated against a cross correlation reference
sub-aperture image that has undergone the same data reduction.  We use FFT-based
cross correlations since they are more efficient than direct cross correlations
when the number of pixels in the correlation field is large. The correlation
function $C$ of sub-aperture $S$ and reference $R$ can be calculated as
\begin{eqnarray*}
  C = {\rm FFT}^{-1} \left[{\rm FFT}[S] \cdot ({\rm FFT}[R])^* \right], &&
\end{eqnarray*}
the asterisk denoting the complex conjugate. 
We apply a $3 \times 3$ pixel 2D-parabolic fit around the brightest pixel to calculate the
position of the correlation maximum $C_{\rm max}$.  The {\it xy}-shift $S_{\rm shift}$
is the error signal of the WFS:
\begin{eqnarray*}
  S_{\rm shift} = R_{\rm ref} - S_{\rm ref} + R_{\rm shift} + C_{\rm max} - 64/2,
\end{eqnarray*} 
$R_{\rm ref}$ and $S_{\rm ref}$ being the spot positions of the perfect
wave-front of cross correlation reference image $R$ and sub-aperture $S$. These are obtained by moving a
pinhole into the entrance focus of the WFS and measuring the pinhole image
positions on the SH detector. 

Since the solar structures evolve, the cross correlation reference image has to be updated regularly. 
The update rate must be shorter than the time scale of the evolution of solar structures
resolved by a sub-aperture, and has therefore been set to 0.2 Hz (every 5~s).  

The {\it xy}-shift $R_{\rm shift}$ of the cross correlation reference image
propagates in the same way
\begin{eqnarray*}
  R_{\rm shift, new} = R_{\rm shift,old} + C_{\rm max} - 64/2.
\end{eqnarray*}

\subsection{Wavefront Reconstruction}
The aberrations that need to be sensed determine the number of Shack-Hartmann sub-apertures: 
\begin{itemize}
\item tip-tilt: needs to be corrected in closed loop
\item defocus (caused by M2 or M3/M4 axial misalignment): needs to be corrected in closed loop (focus drifts due to temperature changes)
\item coma (caused by M2 lateral misalignment): In principle, coma could have been removed in closed loop by a 
lateral movement of M2. However, since a lateral movement of M2 produces 400 waves of tip-tilt for each wave of coma\footnote{One wave (500nm) rms of tip tilt 
aberration corresponds to 0.37 arcsec of image shift.}, a closed-loop coma-correction 
would have had a negative impact on the tip-tilt performance. Instead, the coma was corrected prior to the flight. Since the initial alignment of the telescope was very good, 
only slight lateral M2 adjustments had to be made, so that no vignetting of the F2 field stop by the F1 field stop occured. During the flight, coma was measured only. 
\item spherical aberration (SA, caused by M2 axial misalignment): Although it is in principle possible to remove both defocus and SA simultaneously with M2 and M3/M4, an 
analysis of the optical errors showed that the amount of spherical aberration induced by a defocus of M2 is negligible. 
Therefore it was sufficient to statically align M2 and M3/M4 prior to the flight and correct the in-flight defocus with M2 only. SA does not need to be sensed. 
\item other higher order errors (such as triangular coma due to the M1 support): Interferometric measurements have shown that they have only minor contributions to the 
overall wavefront error, furthermore, there is no degree of freedom for correction anyway. Higher order modes need not to be sensed.  
\end{itemize}

We concluded that a six sub-aperture WFS is sufficient to sense tip-tilt, defocus and coma.  

For sensing these five modes, a simple reconstruction is
sufficient: The reconstruction matrix for the tip-tilt mirror is the
SVD-inverted measured interaction matrix between xy-shifts and the actuators.
The tip-tilt servo is either of PI (proportional + integral) or PID
(proportional + integral + differential) type. The reconstruction matrix for M2
has been pre-calculated via Zernike polynomials. Since the focus drift due to
changing zenith angles occurs on timescales of minutes, M2 receives a
time-averaged (over 10~s, 17000~frames) signal, where a pure proportional
controller is sufficient.  During the flight, only tip-tilt and focus have been corrected in
closed loop. 

\subsection{Real Time Control}
\begin{figure}
  \resizebox{\hsize}{!}{\includegraphics{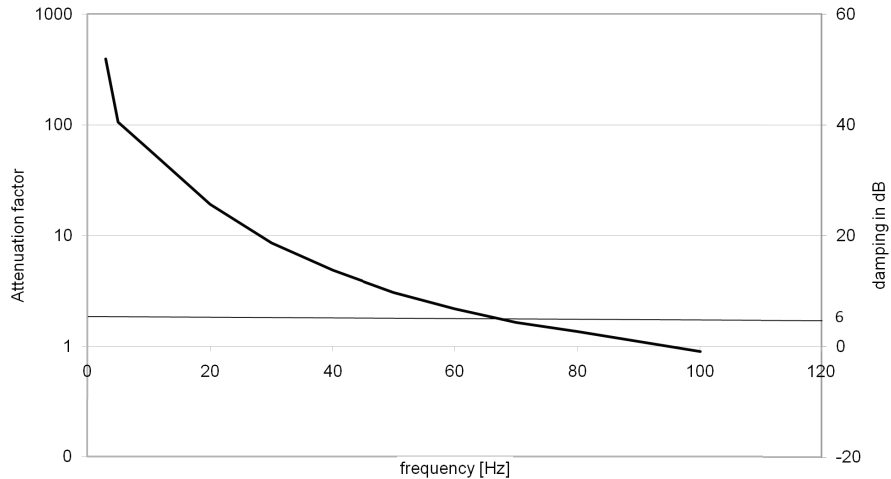}}
  \caption{Performance of the tip-tilt correction system, as measured in the
    laboratory. The 6~dB point, corresponding to a correction of a factor of
    two, is marked as a horizontal line. The corresponding closed-loop frequency
    is 60~Hz.}.
  \label{im:att}
\end{figure}

In order to achieve a high tip-tilt system bandwidth, the delays in the control
loop have to be as small as possible. Listed below are the delays of the
correction:
\begin {itemize}
\item 290~$\mu$s = 1/2 of the inverse of the frame rate 
\item 130~$\mu$s = readout time after which the first two sub-aperture images have been transferred 
  into the frame buffer of the real time computer, and the calculation starts. The total readout time is 380~$\mu$s. 
\item 570~$\mu$s = calculation time\\
  \phantom{570~$\mu$s = }(190~$\mu$s per sub-aperture per CPU, 2~CPUs)  
\item 120~$\mu$s = output + DA conversion  
\item 200~$\mu$s = HV rise time
\item 200~$\mu$s = tip-tilt mirror settling time
\end{itemize}
An additional 290~$\mu$s sample and hold has to be added to the total delay of ca. 1.5~ms between the occurrence of a disturbance and its correction.     
This results in a closed loop 6~dB bandwidth of 60~Hz
for the PID controller and 35~Hz for the PI controller, overachieving the 30~Hz bandwidth specification. 
Figure~\ref{im:att} shows
the disturbance attenuation as a function of the frequency for the PID
controller.
    
The real time control loop software is a variant of the KAOS-system that is used
at the Vacuum Tower Telescope on Tenerife (Berkefeld, 2006).
In order to achieve a compute time of only 190~$\mu$s per sub-aperture
per CPU for a $64 \times 64$ pixel cross correlation field, the compute intensive parts of the code (data reduction and
cross correlation) had to be hand-optimized for the Altivec vector processing
unit of the Motorola G4 CPUs. 

\subsection{Measurement Accuracy}
The measurement accuracy of a correlating SH-sensor depends on the following
properties:
\begin{itemize}
\item image contrast and its (spatial) power spectrum (signal)
\item sub-aperture diameter
\item observing / sensing wavelength
\item pixel scale
\item number of pixels per cross correlation field
\item noise (typically shot noise)
\end{itemize}
The {\it Sunrise} SH-sensor works at 500~nm and uses telescope sub-apertures of 0.33~m,
yielding a spatial resolution of 0.4''. In order to take advantage of this high
(for a SH-sensor) resolution, we use a pixel scale of 0.2''/pixel and
64$\times$64 pixels for the cross correlation (12.8'' field of view).  For
calculating the rms intensity contrast as seen by the wave-front sensor, the original power
spectrum of the solar granulation has to be multiplied by the modulation
transfer function of a sub-aperture. This reduces the original 8.5\% rms granulation
contrast, as measured with the \Imax{}-instrument, to 6\%, as seen by the SH-sensor.

Michau (Michau {\it et al.}, 1992) estimated that the measurement noise variance
$\sigma^2_{\rm snr}$ (in $\lambda^2$) for a critically sampled, correlating
single aperture is
\begin{equation}
  \sigma^2_{\rm snr} = \frac{20\sigma_{\rm noise}^2}{n^2\sigma^2_{\rm signal}}, \label{eq:acc}
\end{equation}
where $\sigma_{\rm noise}$ denotes the rms noise level, $n$ the number of pixels
across the cross correlation field of view (64) and $\sigma_{\rm signal}$ the
image contrast.  The noise level has two main contributors: for the photon noise
we assume a typical use of 160000~e$^-$ (photo electrons) per pixel (80\% full
well capacity), yielding 400~e$^-$.  The 8-bit digitization leads to a noise of
200000~e$^-$/256~=~800~e$^-$ peak-to-peak which corresponds to about
270~e$^-$~rms.  The total noise level is therefore 480~e$^-$.  After applying
Equation (\ref{eq:acc}) and subsequent conversion from a wave-front error (in
$\lambda$) to a pointing (tip-tilt error, in arcsec), the tip-tilt measurement
error for a single 33~cm sub-aperture is $\sigma_{\rm tip-tilt, single} =
0.005''$~rms (per axis).  Averaging over six sub-apertures finally results in
\begin{equation}
  \sigma_{\rm tip-tilt, total} = 0.002''\,{\rm rms}.
\end{equation}

\section {Hardware}
\label{sec:hardware}
The CWS electronics consists of two main parts. The main electronics unit (MEU)
contains the control processor and the processor for the wave-front correction,
as well as all communication and thermal control hardware. The MEU is located
outside the telescope, in the shadow of the solar panels, in order to minimize
any thermal disturbance to the light path, and in order to avoid external
heating by sunlight. The proximity electronics box (PEB) drives the high-speed
tip-tilt mirror and is therefore located close to that mirror.

\begin{figure}
  \resizebox{\hsize}{!}{\includegraphics{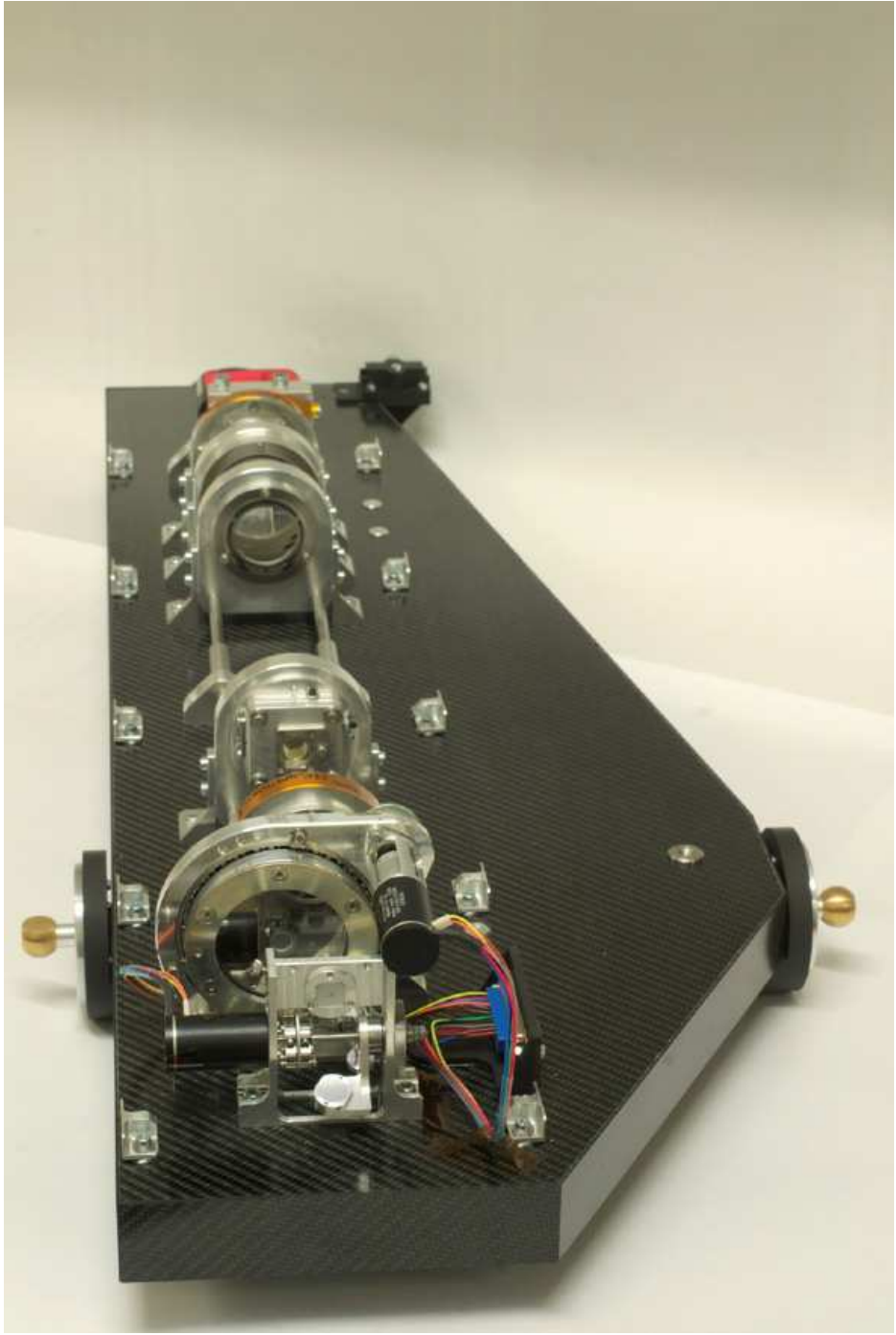}}
  \caption{The wave-front sensor, mounted on its light-weight carbon fiber
    sandwich plate. On the sides, two fixation points of the isostatic mounting are
    visible.}
  \label{cwscompl}
\end{figure}

In order to qualify the hardware for the the near vacuum conditions of the stratosphere, we tested the complete CWS (see Figure~\ref{cwscompl}) and its electronics
at the Instituto Nacional de T\'ecnica Aeroespacial (INTA), Madrid, Spain, at pressures down to 100--200 Pascal and 
temperatures between $+40^{\circ}$ and $-50^{\circ}$ Celsius. No problems were found during these tests.  

\subsection{Thermal and Environmental Design}
At flight altitude the atmospheric pressure is about 5~mbar. To dissipate the
power effectively by thermal radiation, the cooling elements have an allocated
radiation area, and the cooling design has to ensure that the heat is
transmitted to that area. For the CW-electronics-unit we decided to work without
a pressurized box, and to dissipate the power only by conduction and radiation.
All electronics were designed for low power consumption and small power density
to decrease or even avoid any \emph{hot spots}. The advantages of a
conduction cooled electronics unit in contrast to a pressurized box are 
reduced weight and an uncomplicated temperature control with heaters and
radiation coolers. Furthermore, a pressurized box and electronics designed to work under such pressure pose the risk of failure due to loss of air pressure. 
To protect all components from environmental influences, we
applied an electrical isolation and hydrophobic coating to all boards. The
cooling elements had a Lord Aeroglaze A 276 coating with high emissivity and low absorption. The
power consumption of the CW electronics unit in the operating mode is nearly
70~W, 45 of which are used by the real time computer doing the wave-front
correction.

\begin{figure}
  \resizebox{\hsize}{!}{\includegraphics{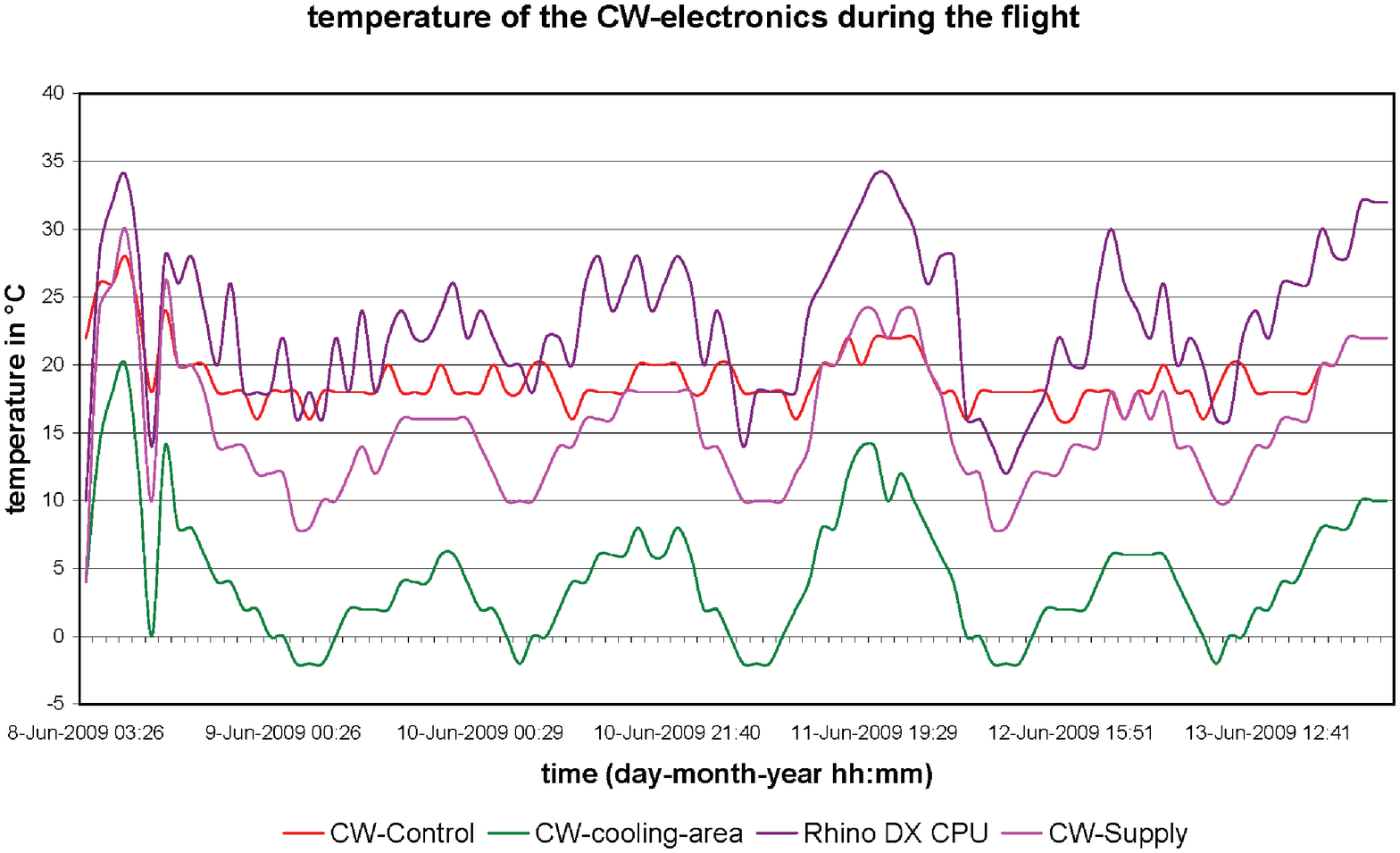}}
  \caption{Temperature profile of the Main Electronics Box during the balloon
    flight. The slow variations correspond to temperature changes caused by the varying elevation of the Sun. The strong peak on 11 June occured during the passage over Greenland, where sunlight was reflected toward the instruments. }
  \label{f:temp} 
\end{figure}

\begin{figure}
  \resizebox{\hsize}{!}{\includegraphics{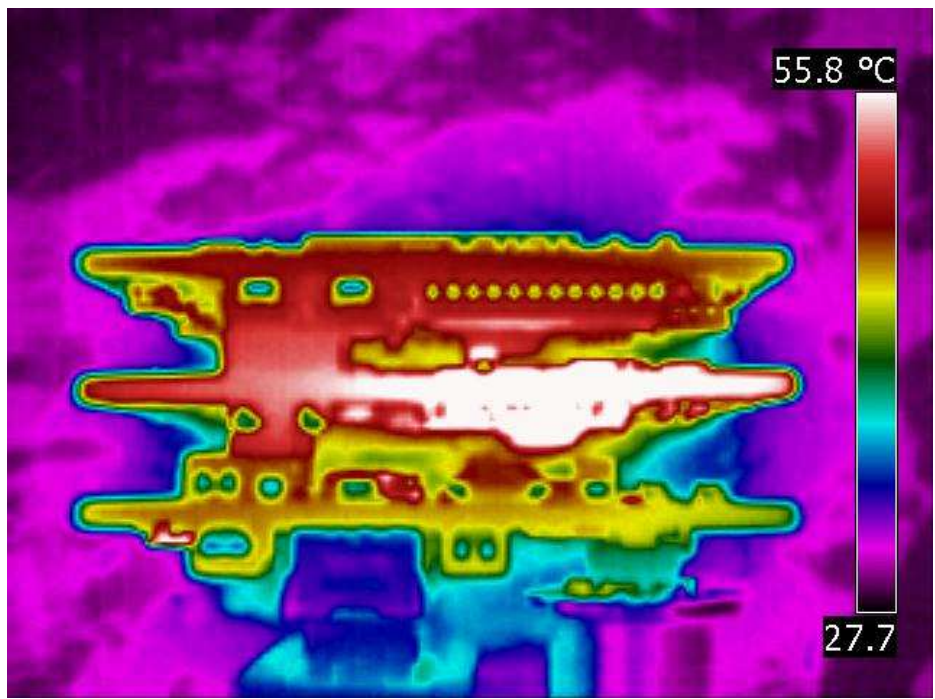}}
  \caption{Infrared image of the electronics boards of the high-speed CMOS camera, prior
    to the installation of the conductive cooling elements. A hot spot is
    clearly visible.}
  \label{f:IR} 
\end{figure}

The rapid fluctuations in the temperature profile in Figure~\ref{f:temp} show the
varying power consumption during the operating and non-operating phases. The
low-frequency pattern shows temperature changes between day and night. The
decreasing temperature at the beginning results from the ascent of the balloon
and the passage through cold atmospheric layers. The noticeable peak on June,
11th coincides with the passage of the gondola over Greenland. The high albedo of the ice led 
to an increase of the temperature. One of the
components of the CWS that required special treatment in order to operate
properly at float altitude, was the high-speed camera. We designed a cooling
element that connected the hot spot in the interior to the walls of the housing.
Figure~\ref{f:IR} shows an infrared image of the camera electronics before the
installation of the cooling element.

\subsection{Main Electronics Unit}
\label{sec:meu}
\begin{figure}
  \resizebox{\hsize}{!}{\includegraphics{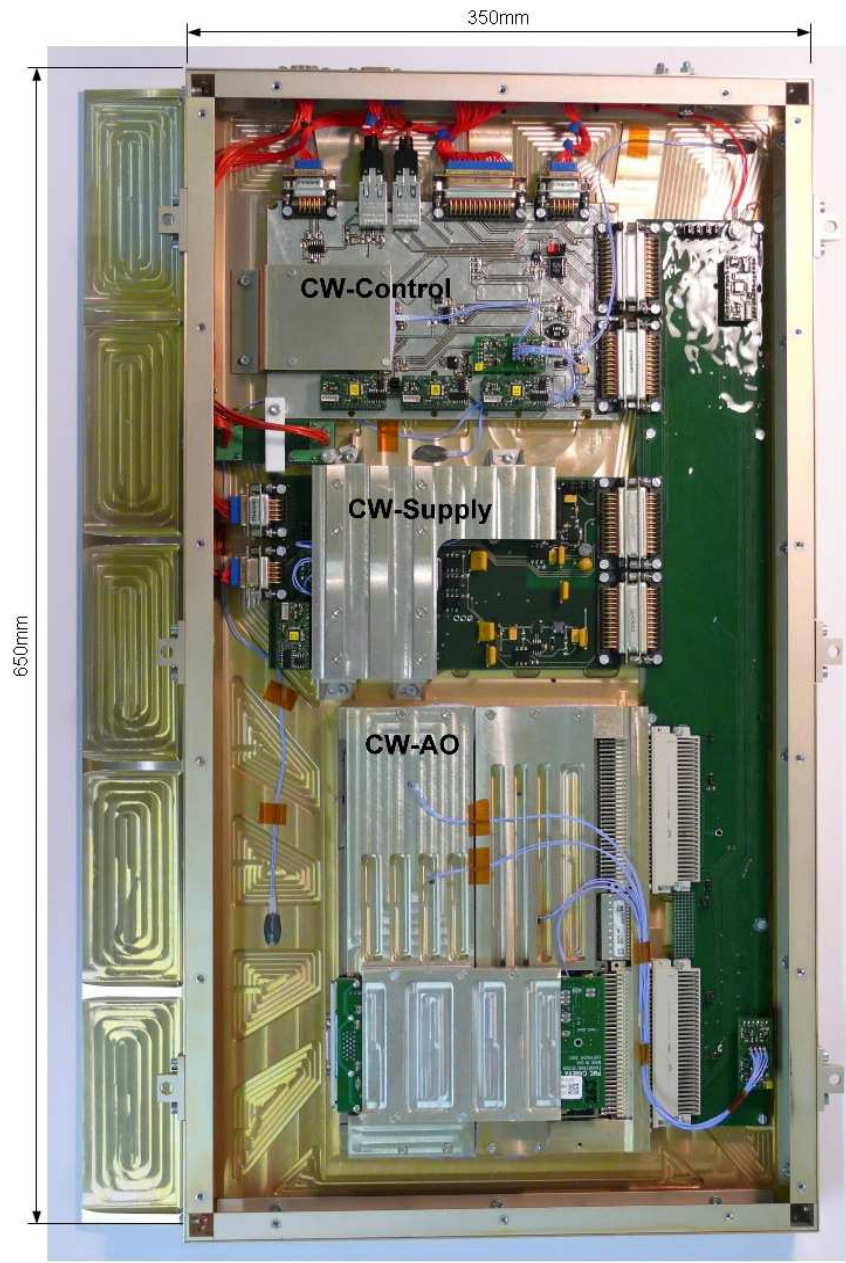}}
  \caption{Main electronics Box of the CWS. `CW Control' denotes the
    control processor (below the cooling plate), the main power supply is
    labeled `CW-Supply', `\Cwao{}' is the control loop real time
    computer (dual Motorola 7457 CPUs)}
  \label{f:meb}
\end{figure}

The CWS main electronics unit (Figure~\ref{f:meb}) contains three different components. The
CW-Control (aka: \Cwcom{}) with its A9M9750 module based on a NetSiliconÕs,
200~MHz, NS9750 micro-controller is a compact low energy solution from
DIGI. It dissipates nearly 1.7~W in operating mode. The CW-Control includes
also the clients of the \Cosm{}-Bus (Volkmer {\it et al.} 2003) This bus handles the control data of the
motion units.  The \Cwao{} is the real time computer that handles the wavefront
correction control loop.  It consists of the Rhino DX Board, a conduction-cooled
VME single board computer with two 1~GHz Motorola 7457 processors, made by
Curtiss Wright, the frame grabber and an interface board, which is specifically
designed for the Rhino DX requirements. Attached to \Cwao{} by two serial
interfaces are the digital-to-analog converters of the tip-tilt mirror.  Also
serially attached to \Cwao{} are the mechanisms that allow changing between dark
stop, field stop and pinhole at the F2 and F6 foci, and the neutral density
filter that keeps the WFS camera illuminated appropriately.  The power
for all components of the CWS is provided from the CW-Supply that has 5
voltage levels with different requirements on currrent or
ripple. The power management is controlled by the
CW-Control, it permits start up and power down of the whole system in a given
order.

\subsection{Proximity Electronics Box}
\begin{figure}
  \resizebox{\hsize}{!}{\includegraphics{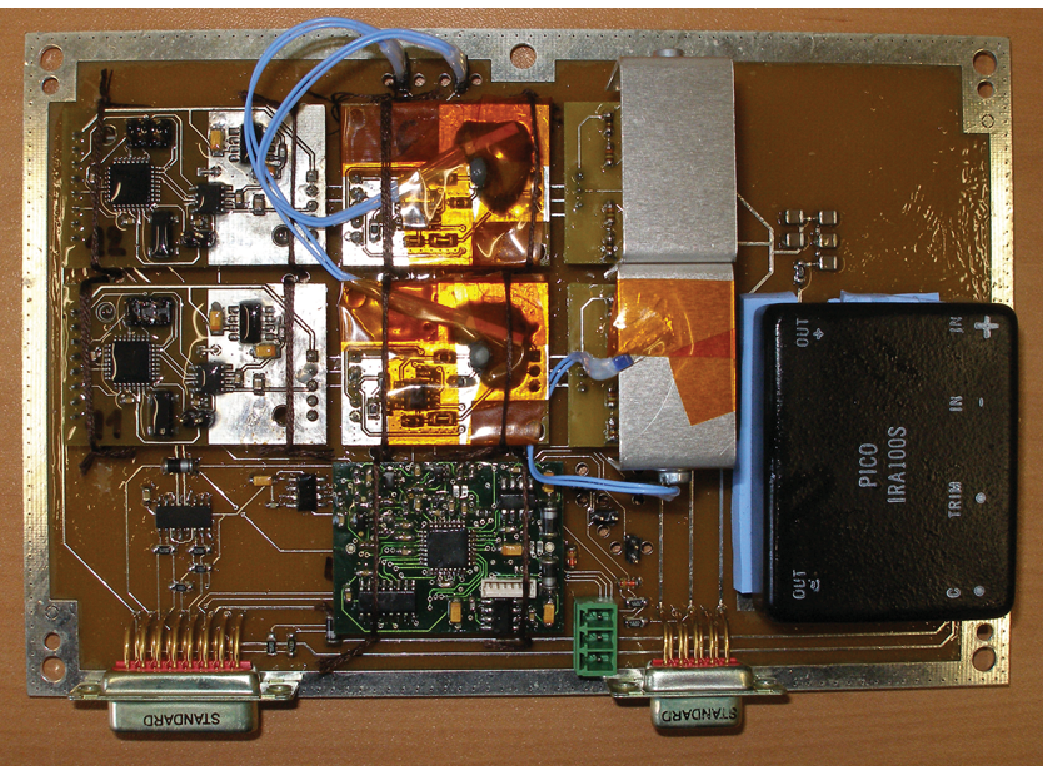}}
  \caption{High-voltage amplifier for the piezo drive of the tip-tilt mirror.
    From left to right: serial interface with digital/analog converter,
    pre-amplifier, main amplifier. The green board at bottom is a \Cosm{} sensor
    server used for temperature measurements. The black box is the 100~V power
    supply for the main amplifier.}
  \label{f:board}
\end{figure}

The tip-tilt mirror (M105) is driven by the high-voltage amplifier of the
proximity electronics box (PEB), see Figure~\ref{f:board}. It requires two variable channels and one
fixed voltage. Each variable amplifier channel consists of a DA
converter with serial interface, a pre-amplifier and a push-pull power stage.
The \Cwao{} dual processor board transmits the data to the high-voltage amplifier
via two RS422 serial interfaces.

\begin{table}
  \caption{CWS Proximity Electronics} 
  \label{table:proxbox}         
  \centering                    
  \begin{tabular}{l l}          
    \hline\hline                
    Component & Characteristics  \\ 
    \hline                      
    
    M105 channel 1		&	7~V...93~V          			\\  
    M105 channel 2		&	7~V...93~V				\\  
    M105 channel 3		&	100~V fixed				\\  
    High voltage supply		&	18~V...36~V $\rightarrow$ 100~V (max. 10~W)\\  
    Push-pull power stage	&	max. 2 * 5~W @ 60~Hz full scale driven	\\  
    Pre-amplifier		&	0~V...5~V $\rightarrow$ 0~V...100~V	\\  
    DA converter interface	&	230.4~kbit s$^{-1}$ 8 N 1 UART			\\  
    DA converter update	        &	max. 10~kHz				\\  
    DA converter resolution	&	16~bit					\\  
    High voltage scale		&	97.6~V / 65535 = 1.489~mV / LSB		\\  
    Dynamic amplifier range	&	86~V					\\  
    Tip-tilt range (on sky)	&	99~arcsec				\\  
    Tip-tilt resolution (on sky)	&	0.00173~arcsec 				\\  	
    \hline                      
  \end{tabular}
\end{table}

The motor controller and the high-voltage amplifier are assembled in the CWS
PEB that is cooled by conduction. All electronic parts with
power losses higher than 250~mW are treated with \emph{thermal filler} and
thereby coupled to the housing. A radiator outside the PEB removes the heat
produced by the CWS proximity box. During the flight, the temperatures in the
box were between -10 $^\circ$C and +20 $^\circ$C.
    
\subsection{Tip-Tilt Mirror}
\label{sec:tt}
\begin{figure}
  \resizebox{\hsize}{!}{\includegraphics{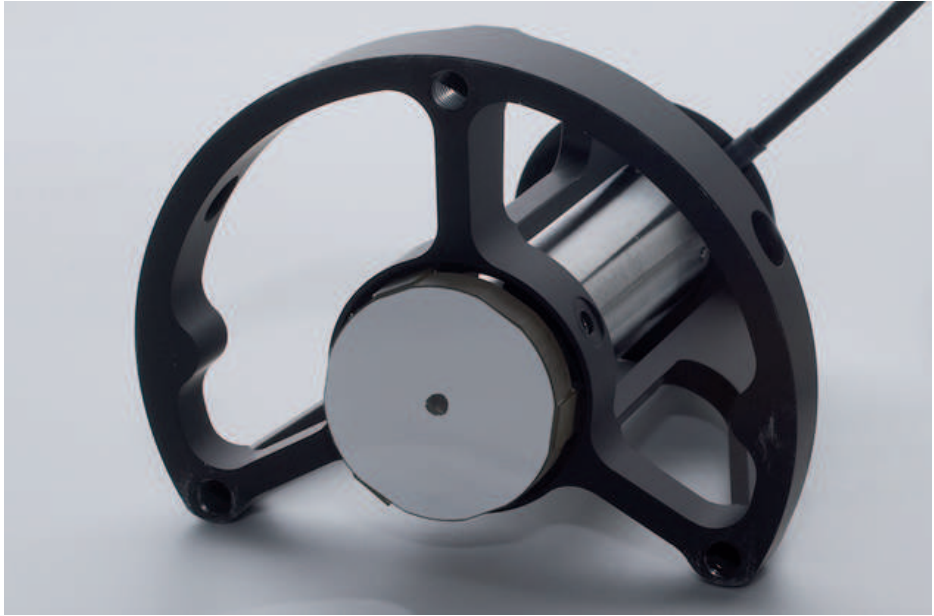}}
  \caption{Tip-tilt mirror with the piezo stage (in the polished cylinder) and
    the mechanical interface to the mount assembly.}
  \label{TT}
\end{figure}

\begin{figure}
  \resizebox{\hsize}{!}{\includegraphics{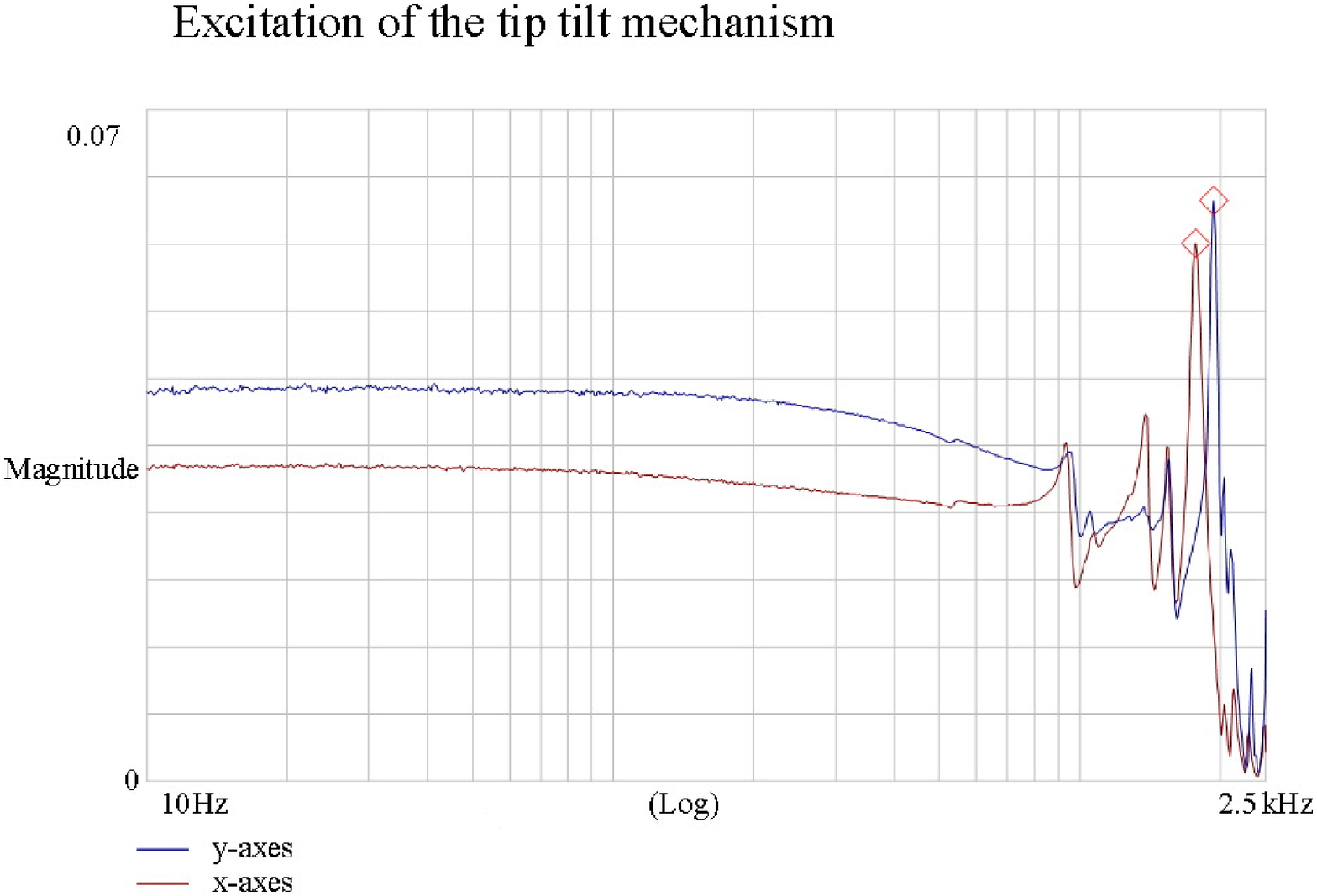}}
  \caption{Excitation of the tip-tilt mirror as a function of the sinusoidal
    excitation frequency.}
  \label{TTexcitation}
\end{figure}

For the tip-tilt correction a two axis piezo ceramic actuator from \emph{Physik Instrumente} was
used (P.I. S-330K065). 
This  piezo
drive fulfills all requirements for the {\it Sunrise} image stabilization
(high acceleration, fast response to set points and small increments). The 
piezo actuators have a total tilt range of 10~mrad (corresponding to
$\pm$~45~arcsec in the sky). The minimal step size corresponds to 0.15~$\mu$rad
(1.7~mas on the sky).  The dynamic characteristics are also shown in
Table~\ref{table:proxbox}. The piezo actuators have a very low power
dissipation. 
The zerodur tip-tilt mirror attached to the actuator is 35~mm in diameter with a thickness of
7.5~mm (see Figure~\ref{TT}). The surface flatness (75.9~nm peak to valley and 12.6~nm rms after attaching the piezos) is
important because of its location close to the pupil image. Therefore, it is
necessary to mount the mirror in a way that decouples the optical surface from
differential thermal expansion of the mechanical parts and to reduce the
mounting force as much as possible.  The eigenfrequency of the whole system
(mirror actuator and mounting) is about 1900~Hz (see Figure~\ref{TTexcitation}). At
1700~Hz closed loop frequency the drive is operated in a minimum of excitation.

\subsection{Wave-Front Sensor Camera}
\label{sec:camera}
The WFS detector is an industrial DALSA 1M150 CMOS camera that has the following
properties and settings :
\begin{itemize}
\item 8~bits per pixel
\item full well capacity = 200~ke$^-$
\item fill factor $\times$ quantum efficiency @ 500~nm = 15\%
\item 1024$\times$1024 pixels
\item region of interest (ROI) = 256$\times$224 pixels
\item framerate(ROI) = 1700~Hz
\item integration time = 100~$\mu$s 
\item readout time = 380~$\mu$s
\item power consumption = 2~W
\end{itemize}
Since the camera is exposed to ca 80\% of the full well capacity, the noise
level (in e$^{-}$) is dominated by shot noise and digitization noise is slightly less important.  
Readout noise is negligible. Thermal filler was inserted between the electronic
components of the camera in order to guarantee conduction cooling at
float altitude.

\section{Control Software}
\label{sec:sw}

This section deals with the CWS flight control software as run on the embedded
ARM micro-computer called \Cwcom{} described in Section~\ref{sec:meu}. The
software communicates with the Instrument Control Unit (ICU), accepting commands either originating from
the ICU itself or from the CWS Electronic Ground Support Equipment (EGSE). In return housekeeping data is sent to the
ICU. The \Cwcom{} also controls the behavior of the \Cwao{} computer ({\it cf.}  Section~\ref{sec:meu}) by sending commands to it. Status
information of the \Cwao{} computer is retrieved and forwarded as housekeeping to
the ICU.  The communication protocol between \Cwcom{} and \Cwao{} is the same as at
the GREGOR ground-based solar telescope (Volkmer {\it et al.}, 2006). Furthermore \Cwcom{} communicates with the main telescope
controller (MTC) and the pointing system (PS).

Section~\ref{sec:cwcom-states} describes the state-driven operation of the
CWS as used during flight. Thereafter the behavior of the CWS in each specific
state is discussed in Section~\ref{sec:states} while
Section~\ref{sec:cwcom-mtc} gives an overview of the communication between
\Cwcom{} and the MTC.

\subsection{State Driven Operation of \Cwcom{}}
\label{sec:cwcom-states}

Due to the software design CWS is a state driven instrument.  The different
states, the possible state transitions, and the behavior of the software within
each state are described below.

\begin{figure}[h!]
  \centering
  \includegraphics[width=0.45\textwidth]{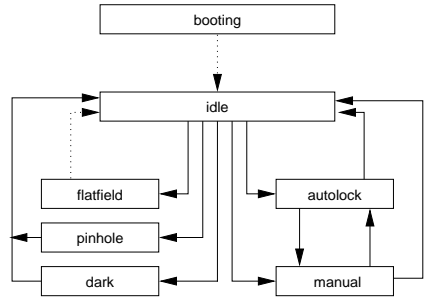}     
  \caption{CWS state chart.}
  \label{fig:cws-statechart}
\end{figure}

Figure~\ref{fig:cws-statechart} illustrates the transitions between the
seven possible states which are discussed in more detail below. Arrows with solid
lines denote transitions that require the \texttt{setstate} command (either from
the ICU or from an EGSE) while arrows with dotted lines represent state changes
that are done automatically.

\subsection{States}
\label{sec:states}
This section discusses each of the individual states CWS might adopt. The pre-
and postconditions for each state are discussed. Furthermore the behavior within
each state is described here in detail.

\subsubsection{booting}
\label{sec:state-booting}

The state \textbf{booting} is adopted once the CWS is powered up. While the CWS
is in this state it does not accept any command, it will 					
deliberately ignore each command trying to change the state. While the CWS is in
this state, it will power up \Cosm{}, the proximity electronics box, the camera and
the \Cwao{}. It will remain in this state until \Cwao{} responds to the
\texttt{status} command, {\it i.e.}, provides the necessary data that enable the
\Cwcom{} to send reasonable HK-data.
 

\subsubsection{idle}
\label{sec:state-idle}

This state is intended for manual operation of the CWS. The instrument will not
take any action in this state except sending HK data.  Whenever the idle state
is reached, \Cwcom{} will ramp the tip-tilt mirror and will make sure that the AO
loop is open and F2 is set to field stop (light passes to the instruments). The
CWS will accept any command including state transition commands in this state.

\subsubsection{flatfield}
\label{sec:state-flatfield}

In order to initiate a flatfield calibration, the ICU commands the CWS to the
\textbf{flatfield} state. Although this state is called {\it flatfield}, it performs additional tasks. 
The CWS flatfielding procedure takes about 6~minutes and consists of the
following steps:

\begin{enumerate}
\item ensure AO-loop is open
\item adjustment of filter wheel for optimal light level
\item take the actual flatfield
\item set the F6 position to pinhole
\item mark and measure the F6 reference spot positions of the Shack-Hartmann sensor that define the perfect wavefront 
\item set F6 position to field
\end{enumerate}

The CWS does not accept state transition commands while it is in this state and
will reach the \textbf{idle} state after completion of flatfielding.

\subsubsection{pinhole}
\label{sec:state-pinhole}
Taking F2-pinhole images is also commanded by the ICU. CWS moves the tip-tilt
mirror to its optical zero position and sets the F2 unit to the pinhole position.
The minimum time CWS will remain in this state is 2~s. While in this state, CWS does not
accept state transition commands except to the \textbf{idle} state.

\subsubsection{dark}
\label{sec:state-dark}

If a \texttt{setstate} \textbf{dark} command is received this state is adopted.
\Cwcom{} commands the F2 unit to the dark position 
before taking darks.
The CWS requires about 10~s for
the dark procedure to complete.  It does not accept state transition commands
while in this state and will only accept a transition command to the \textbf{idle} state
after completion of the dark procedure.

\subsubsection{autolock}
\label{sec:state-autolock}
This is the state intended for automated observations. A typical observation
scenario looks like this:

\begin{enumerate}
\item point the telescope to the region of interest
\item take a flatfield (\texttt{setstate flatfield})
\item wait for all instruments to complete their flatfield
\item set CWS to autolock mode (\texttt{setstate autolock})
\item wait for tip-tilt and focus loop to close (see field \texttt{AO\_runmode}
  in HK of CWS)
\item do actual observation sequence
\item open tip-tilt/focus loop by setting CWS to idle (\texttt{setstate idle})
\end{enumerate}

The behavior of the CWS in the \textbf{autolock} state is split into two parts,
a startup phase that persists until a lock (tip-tilt plus focus) is achieved for
the first time for at least 20~s (provided reasonable residuals are
achieved) and an operational phase thereafter.  During the startup phase a new
reference image is taken before each attempt to close the loop.  If the CWS loop
crashes more than five times in a row or a lock signal from the PS is
not received within 150~s during the operational phase ({\it e.g.} due to image motion beyond the range
of the tip-tilt mirror) it will restart with a startup phase.

Note that the CWS will never leave the \textbf{autolock} state unless it
receives a setstate \textbf{idle} or \textbf{manual} command. Especially if the
CWS is not able to get a lock at the position the telescope points to ({\it e.g.} at
the solar limb) it will remain in the startup phase of the autolock state
forever. It is the responsibility of the ICU (or an operator of the EGSE) to
monitor that a lock was achieved within a reasonable time.

\subsection{MTC Communication}
\label{sec:cwcom-mtc}

The main telescope controller (MTC) is physically connected to the \Cwcom{} using
an RS422 interface, hence, all communication between {\it Sunrise} and the telescope
must pass \Cwcom{}.  All communication between the MTC and the ICU is
forwarded by \Cwcom{} transperantly. The only commands sent to the MTC
directly are positioning commands for the M2 mirror for focus and coma correction.



\section{In-Flight Performance}
\label{sec:perf}
\begin{figure}
  \resizebox{\hsize}{!}{\includegraphics{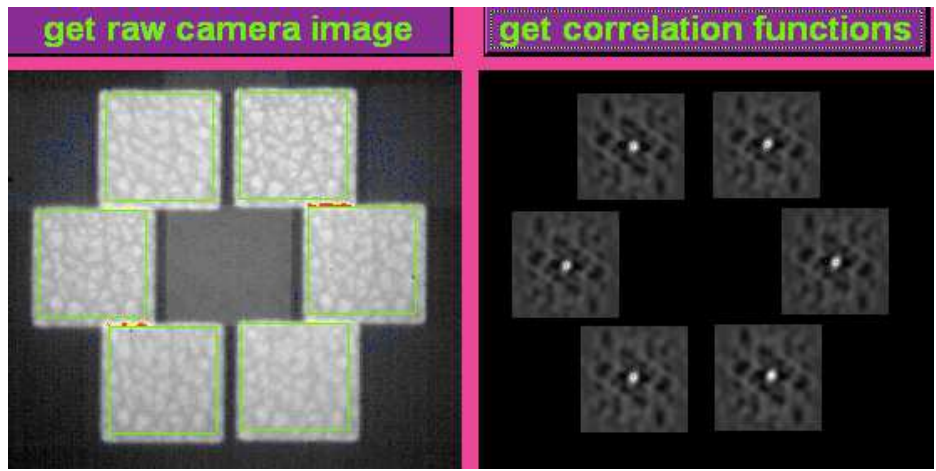}}
  \caption{Sub-images and correlations functions of the wave-front sensor during
    closed-loop tracking. Screen shot taken from the flight-control system.}
  \label{im:imwfs}
\end{figure}

After the launch and opening of {\it Sunrise}'s aperture door, the CWS was the first
instrument to be powered up and put into operation. Figure~\ref{im:imwfs} shows
one of the first WFS images (left) and the corresponding correlation functions.
The excellent quality of both image and correlation functions (compared to a
ground based solar SH-system) is obvious.
\subsection{Times of Closed Loop} 
\begin{figure}
  \resizebox{\hsize}{!}{\includegraphics{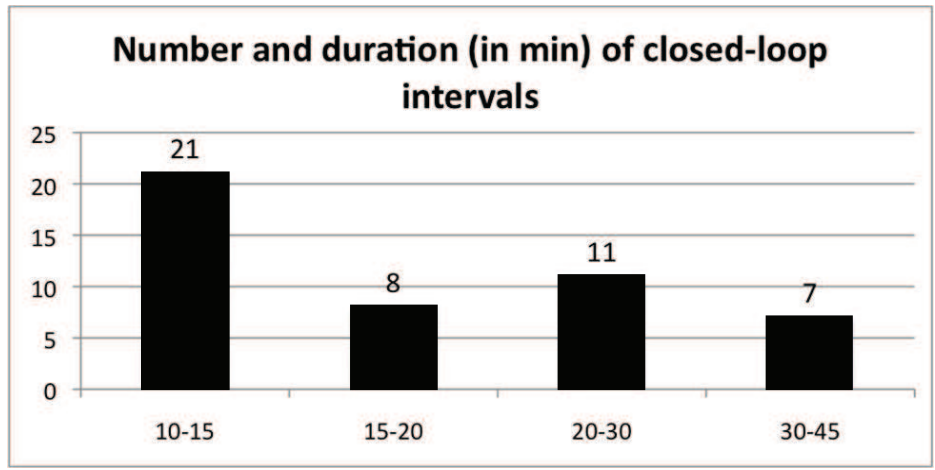}}
  \caption{Number of closed-loop intervals longer than ten minutes. The total
    time with closed-loop observations was 33 hours.}
  \label{im:cl}
\end{figure}

Figure~\ref{im:cl} shows the number of closed loop intervals as a function of the
duration of the intervals. It can be seen that there is no time series longer
than 45~minutes.  This is due to the fact that the gondola had problems 
stabilizing the telescope to within $\pm$45~arcsec (range of the fast tip-tilt
mirror) whenever shear winds occurred.  Between Norway and Greenland, over the
free ocean, wind gusts occurred less often than over Greenland or over the many
coastlines during the last day.  The lack of bandwidth of the gondola pointing system to correct wind gusts is
one of the problems that were found during the flight. The overall observation
time with image stabilization, {\it i.e.} without the observation overhead (flatfielding, calibrations, repointing etc) 
was more than 33~h.

\subsection{Residual Tip-Tilt Error}

\begin{figure}
  \resizebox{110mm}{!}{\includegraphics{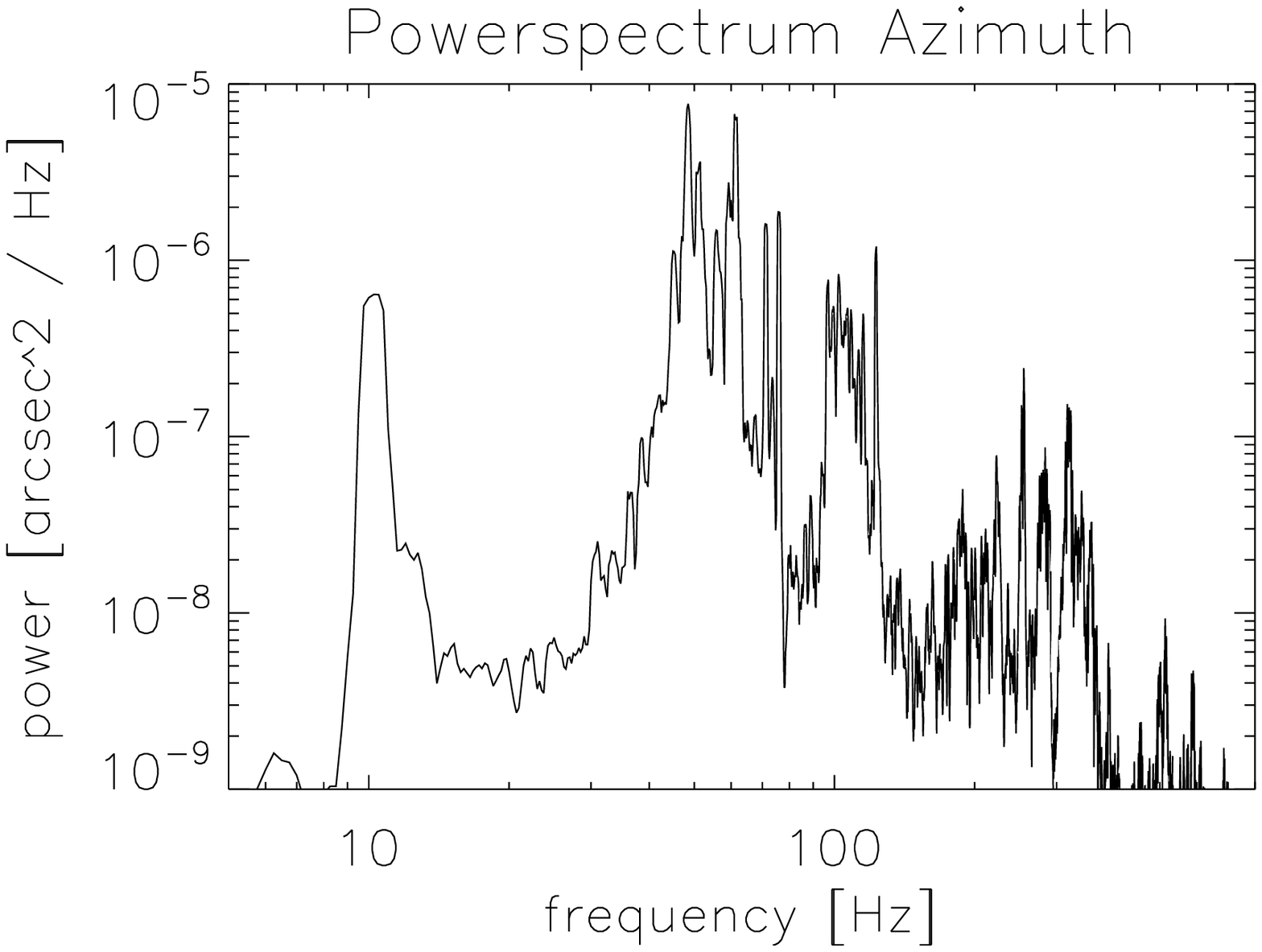}}
  \resizebox{110mm}{!}{\includegraphics{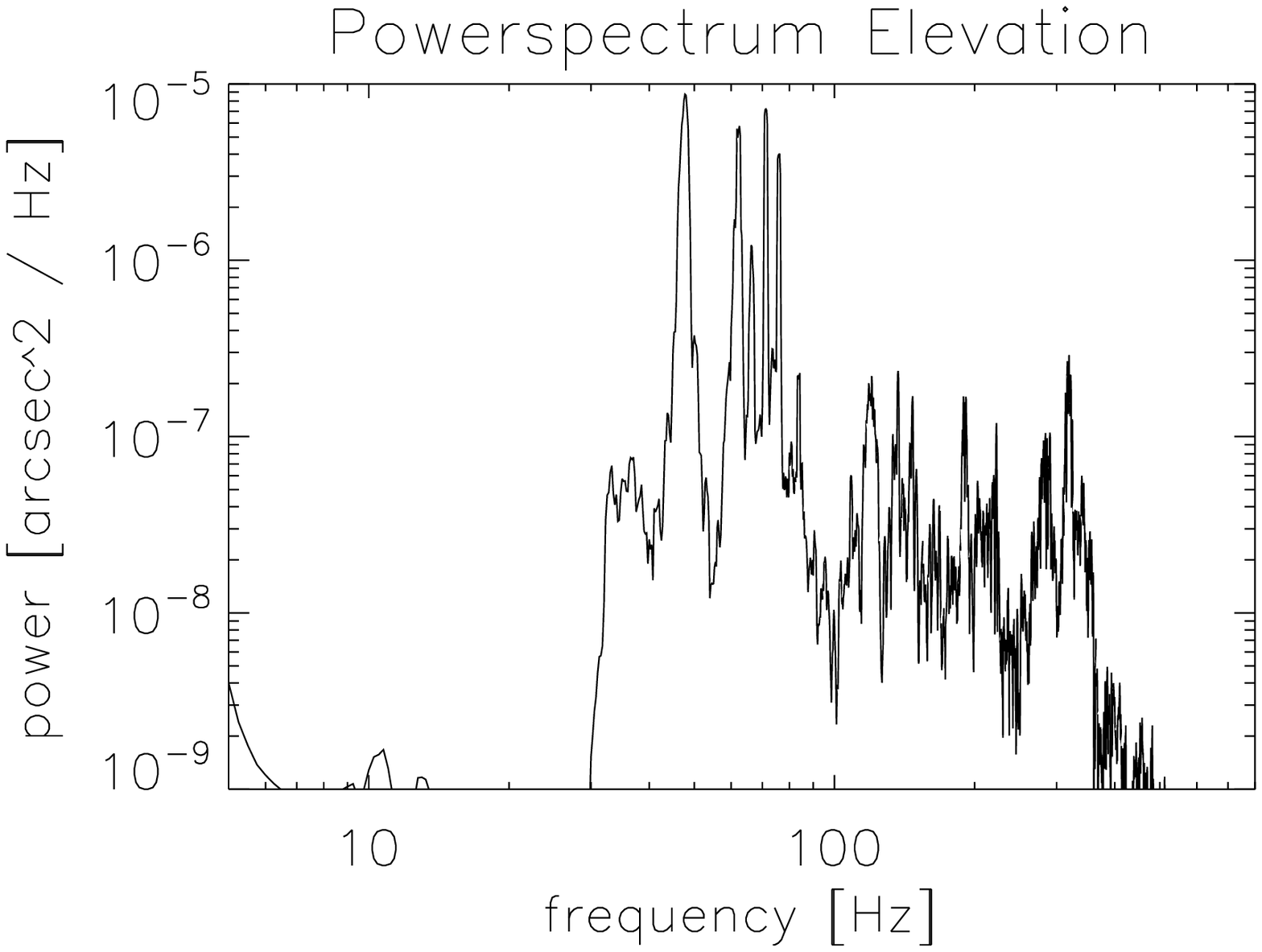}}
  \caption{Power spectra of residual image jitter, for the azimuth (top panel),
    and elevation axis (bottom panel).}
  \label{im:ps}
\end{figure}

A second problem showed up when analyzing the power spectrum of the residual
tip-tilt error. Figure~\ref{im:ps} shows the power spectrum for the azimuth (top)
and elevation (bottom) as a function of the frequency. It can be seen that
vibrations in the frequency range of 30--90~Hz occur that were only partially
damped (see Figure~\ref{im:att}).  Vibrations above 90~Hz are even amplified a
bit. As a consequence, the residual image motion could not be reduced below
0.03-0.04''~rms. The vibrations were most probably caused by the gondola.  The
telescope itself does not transmit frequencies below 30~Hz (with the exception
of 10~Hz), however, frequencies above 30~Hz could propagate from the gondola
through the telescope. As a consequence, short exposure images (say, 30~ms, as used in some SuFI
wavelengths) have the
same image quality as images that were exposed for 30~s (as used by IMaX and SuFI at 214 nm).

For the next flight, vibration damping measures for the gondola have to be
taken. Furthermore, it is possible to increase the tip-tilt bandwidth by about
70\% by using only two SH-sub-apertures instead of six during closed loop,
allowing the detection of image shift and focus only. The resulting measurement
accuracy will be reduced to 0.0035''~rms. Changing between two and six subapertures can be done in flight because it only requires a change of the parameter set. 

\subsection{Focus Correction}
The in-flight focus correction proved to be vital. The uncorrected daily focus
drift was $\pm$0.5$\lambda$ rms between high and low solar elevations.
By moving M2 in closed loop during observations, the focus error was reduced to
0.01$\lambda$~rms at the CWS entrance focus. Phase diversity analysis of the flight data shows a differential focus of $\lambda/20$ between SuFI and CWS 
and of $\lambda/20$ between \Imax{} and CWS, 
well within the range to be corrected by using the phase diversity
capabilities of SuFI and \Imax{}.

\section{Discussion}
\label{sec:concl}
During its successful flight in 2009, the {\it Sunrise} telescope had by far the
highest pointing stability at the scientce focus ever achieved on a balloon-borne telescope. During
its five day flight, it provided scientific data of seeing-free, unprecedented polarimetric 
accuracy and high resolution observations at UV wavelengths never observed before at similar resolution. 

After the vibrations problems of the gondola have been solved, 
the goal of 0.005~arcsec rms pointing stability can be reached with the original six sub-aperture setup of the CWS. 
If some vibrations remain, the pointing could still be improved to ca. 0.01~arcsec rms by using the increased bandwidth of the two sub-aperture setup.
The Sunrise team is therefore aiming for a second flight, ideally in 2012 close to the solar maximum."

\begin{acks}
The German contribution to {\it Sunrise} is funded by the Bundesministerium
f\"{u}r Wirtschaft und Technologie through Deutsches Zentrum f\"{u}r Luft-
und Raumfahrt e.V. (DLR), Grant No. 50~OU~0401, and by the
Innovationsfond of
the President of the Max Planck Society (MPG). The Spanish contribution has
been funded by the Spanish MICINN under projects ESP2006-13030-C06
and AYA2009-14105-C06 (including European FEDER funds). The HAO
contribution was partly funded through NASA grant number NNX08AH38G.
\end{acks}

\end{article} 
\end{document}